\documentclass[11pt]{article}
\usepackage{hyperref}
\usepackage{cite}
\usepackage{footnote}
\makesavenoteenv{minipage}
\renewcommand{\title}[1]{%
    \bigskip%
    \begin{center}%
    \Large\bf #1%
    \end{center}%
    \vskip .2in}

\renewcommand{\author}[1]{%
    {\begin{center}
    #1
    \end{center}}}
\newcommand{\address}[1]{\vspace{-1.7em}\vspace{0pt}
    {\begin{center}
    \it #1
    \end{center}}}

\begin{document}

\begin{titlepage}
\title{New Hamiltonian analysis of Regge Teitelboim minisuperspace cosmology}

\author
{
Rabin Banerjee  $\,^{\rm a,b}$,
Pradip Mukherjee $\,^{\rm c,d}$,
Biswajit Paul    $\,^{\rm a, e}$}
\address{$^{\rm a}$S. N. Bose National Centre 
for Basic Sciences, JD Block, Sector III, Salt Lake City, Kolkata -700 098, India }

\address{$^{\rm c}$Department of Physics, Barasat Government College,\\Barasat, West Bengal

 }

\address{$^{\rm b}$\tt rabin@bose.res.in}
\address{$^{\rm d}$\tt mukhpradip@gmail.com}
\address{$^{\rm e}$\tt bisu\_1729@bose.res.in}
\begin{abstract}

A new Hamiltonian formulation of the minisuperspace cosmology following from the geodetic brane gravity model introduced by  Regge and Teitelboim is presented. The model is considered in the framework of higher derivative theories which facilitates Hamiltonian formulation. The analysis is done using the equivalent first order approach.  The gauge generator containing the exact number of gauge parameters is constructed. Equivalence between the gauge and reparametrization symmetries has been demonstrated. Complete gauge fixed computations have been provided and formal quantization is done indicating the Wheeler de Witt equation. Compatibility with existing results is shown.
\end{abstract}
\end{titlepage} 

\section{Introduction}
Higher derivative(HD) theories were once introduced as a possible mechanism of renormalization. By higher derivative theory we mean those theories with Lagrangian depending on higher order time derivative of the fields than the first. Recently, interest in this field is rekindled due to the advent of higher order theories of gravitation. An interesting occurrence of higher derivative terms in the action appears in General Relativity. There, usually, such terms are isolated as surface terms and dropped. However in case of gravity the surface term is always not ignorable e.g. the requirement of the Gibbons - Hawking term in the action. This is more so in the brane world scenario where the universe is viewed as a hypersurface immersed in a bulk. A classic model is due to Regge and Teitelboim (RT) \cite{regge} where gravitation is described as the world volume swept out by the motion of a three - dimensional brane in a higher dimensional Minkowski spacetime. Hamiltonian analysis of the model and its quantization was further explored in \cite{davidson1,  karasik, cordero1}. Unlike the Einstein gravity, in the RT model the independent fields are the embedding functions rather than the metric. In the  RT model second derivatives of the fields appear in the action and like general relativity these higher derivative terms may be clubbed in a surface term. In the usual formulation this surface term is dropped \cite{karasik} thereby reducing the original model to a first order theory. However this makes the Hamiltonian formulation of the model problematic \cite{karasik}. These problems are bypassed
 by introducing an auxiliary field \cite{karasik}. On the other hand recently it has been pointed out  that no such auxiliary field is needed if one includes the surface term in the RT model containing higher derivative terms \cite{cordero1}. Obviously, therefore, the Hamiltonian formulation of this model is far from closed. The present paper addresses this and related issues.

 Higher derivative theories were studied and used in different contexts over a long period of time \cite{cordero1, podolsky1, podolsky2, Pais, pisarski, nesterenko, plyuschay1, BMP, BPU, elie, Iliopoulos, Gama, gib, caroll, woodard1, neupane, nojiri4, ani, berg1, berg2}.
Though the classical Hamiltonian formulation of higher derivative theories was worked out by Ostrogradsky long ago  \cite{ostro} and has been refined over the years, specifically in the context of gauge theories certain aspects of the Hamiltonian formulation were not adequately emphasised. One such issue is the mismatch between the number of primary first class constraints and the number of independent gauge degrees of freedom in a higher derivative relativistic particle model \cite{nesterenko}. Recently it has been demonstrated \cite{BMP} that under  an equivalent first order formalism \cite{plyuschay1} which is a variant of the Ostrogradsky approach, the well known algorithmic method of construction of the gauge generator for first order systems \cite {BRR1,BRR2} can be invoked to settle the issue. The Hamiltonian method developed in 
\cite {BMP} of abstracting the independent gauge degrees of freedom of higher derivative systems has been applied to a number of particle and field theoretic models \cite{BMP, MP, paul} successfully. Note in this context that the anasysis of the RT model in the ambit of higher derivative theory \cite{cordero1} was done from the Ostrogradsky approach and this work is based on the minisuperspace model following from the RT theory. The minisuperspace model carries the reparametrization invariance of the original RT gravity which appears as gauge invariance in the Hamiltonian analysis. It will naturally be interesting to apply the equivalent first order formalism of \cite{BMP} to the RT model with the surface term. This will be the subject of the present paper. Like \cite{cordero1} the analysis will be based on the minisuperspace model.

   Before finishing the introductory comments it will be appropriate to say a few words about the equivalent first order formalism. This method of treating  higher derivative systems can be distinguished easily from the usual Ostrogradsky approach. In both the approaches successive time derivatives of the coordinates are considered as phase space variables right upto one order less than the highest derivative appearing in the Lagrangian. Corresponding momenta are introduced to complete  the phase space. The relations between the `coordinates' of the enlarged phase space is reflected in the Ostrogradsky method in the choice of momenta which have to be defined in a particular way to account for the higher derivative nature.
In contrast, in the equivalent first order formalism such relations are accommodated as Lagrangian constraints so that momenta are defined in the usual way as is done for the first order theories. This introduces new restrictions on the variations in phase space which is not apparent in the Ostrogradsky method. This difference was instrumental in the construction of the Hamiltonian gauge generator \cite{BMP} that could explain the apparent mismatch in the number of independent gauge degrees of freedom with the number of independent primary first class constraints reported in \cite {nesterenko}. Note that  in \cite {nesterenko} the Ostogradsky approach of Hamiltonian formulation was adopted. The equivalent first order formalism also provides a straightforward Hamiltonian procedure a la Dirac \cite{dirac} to treat the singular systems endowed with gauge symmetry. The analysis of the RT model with the higher derivative terms from the point of view of the equivalent first order formalism is thus interesting in its own right.

 The structure of the paper  is as follows. In section 2 a review of the cosmological model based on RT gravity is provided. This will also help us in fixing notations. In section 3 Hamiltonian formulation of the RT cosmology is discussed. This is a new Hamiltonian formulation of the model which like \cite{cordero1} retains the higher derivative term but, contrary to \cite{karasik}, is based on the equivalent first order formalism of treating higher derivative system rather than the usual Ostrogradsky approach. Analysis of independent gauge symmetries is given which is demonstrated to be consistent with the Lagrangian (reparametrization) invariance  of the model. An exact mapping between the gauge and  reparametrization parameter has been worked out. Gauge fixing has been done and an appropriate symplectic algebra in the form of the Dirac brackets between the phase space variables has been given. Using the strongly implemented(second class) constraints the phase space is reduced and the number of independent phase space variables is found to be two. 
Finally formal quantization is indicated in the usual way \cite {hanson}. The Wheeler DeWitt(WDW) equation is constructed in the fully reduced phase space. Its compatibility with the results existing in the literature \cite{davidson1} is demonstrated. Our conclusions are given in section 4.


\section{Regge--Teitelboim cosmogical model}
The RT model considers a d-dimensional brane $\Sigma$ which evolves in a $N$ dimensional bulk spacetime with fixed Minkowski metric $\eta_{\mu\nu}$. The world volume swept out by the brane is a $d+1$ dimensional manifold $m$ defined by the embedding $x^\mu = X^\mu(\xi^a)$ where $x^\mu$ are the local coordinates of the background spacetime and $\xi^a$ are local coordinates for $m$. The theory is given by the action functional
\begin{equation}
S[X] = \int_{m} d^{d+1} \xi  \sqrt{-g} (\frac{\beta}{2}\mathcal{R} - \Lambda), \label{main_lag}
\end{equation}
where $\beta$ has the dimension $[L]^{1-d}$ and $g$ is the determinant of the induced metric $g_{ab}$. $\Lambda$ denotes cosmological constant and  $\mathcal{R}$ is the Ricci scalar. As has been already stated above, we will be confined to the minisuperspace cosmological model following from the RT model. 

 The standard procedure in cosmology is to assume that on the large scale the universe is homogeneous and isotropic. These special symmetries enable the $4$ dimensional world volume representing the evolving universe to be embedded in a 5-dimensional Minkowski space time 
 \begin{equation}
 ds^{2} = - dt^{2} + da^{2} + a^{2}d\Omega_{3}^{2},
\end{equation}  
where $d\Omega_{3}^{2}$ is the metric for unit 3 sphere. To ensure the FRW case we take the following parametric representation for the brane
\begin{eqnarray}
x^{\mu} = X^{\mu}(\xi^{a}) = \left( t(\tau),a(\tau), \chi, \theta, \phi \right),
\end{eqnarray}
 $a(\tau)$ is known as the scale factor.

 After ADM decomposition  with space like unit normals ($N=\sqrt{\dot{t}^{2}-\dot{a}^{2}}$ is the lapse function)
 \begin{equation}
 n_{\mu} = \frac{1}{N}(-\dot{a}, \dot{t}, 0,  0,0),
 \end{equation}
   the induced metric on the world volume is given by,
  \begin{equation}
  ds^{2} = -N^{2} d\tau^{2} + a^{2} d \Omega_{3}^{2}.
\end{equation}
Now, one can compute the Ricci scalar which is given by
\begin{equation}
\mathcal{R} = \frac{6 \dot{t}}{a^{2} N^{4}}(a \ddot{a}\dot{t} - a\dot{a}\ddot{t} + N^{2}\dot{t}).
\end{equation}
With these functions we can easily construct the Lagrangian density as
\begin{equation}
\mathcal{L} = \sqrt{-g} \left(\frac{\beta}{2}\mathcal{R} - \Lambda\right).
\end{equation} 
The Lagrangian in terms of arbitrary parameter $\tau$ can be written as\cite{cordero1}\footnote{here $H^2=\frac{\Lambda}{3\beta}$, a constant quantity}
\begin{equation}
L(a, \dot{a}, \ddot{a}, \dot{t}, \ddot{t}) = \frac{a\dot{t}}{N^{3}} \left({a \ddot{a} \dot{t}-a \dot{a} \ddot{t} + N^{2}\dot{t}} \right) - N a^{3} H^{2}.
\label{hdlag}
\end{equation}
Varying the action with respect to the field $a(\tau)$ we get 
 the corresponding Euler Lagrange equation as 
\begin{equation}
\frac{d}{d \tau}\left( \frac{\dot{a}}{\dot{t}} \right) = - \frac{N^{2}}{a \dot{t}}\frac{(\dot{t}^{2}- 3N^{2}a^{2}H^{2})}{(3\dot{t}^{2}- N^{2}a^{2}H^{2})}.
\label{ELeqn}
\end{equation} 
Note that the Lagrangian (\ref{hdlag}) contains higher derivative terms of the field $a$. However we can write it as \cite{cordero1}
\begin{equation}
L= -\frac{a{\dot {a}}^2}{N} + aN\left(1 - a^2H^2\right) + \frac{d}{d\tau}\left(\frac{a^2{\dot{a}}}{N}\right).
\label{orglag}
\end{equation}
If we neglect the boundary term the resulting Lagrangian becomes usual first order one. As is well known the equation of motion is still given by (\ref{ELeqn}). However the Hamiltonian analysis is facilitated if we retain the higher derivative term. Thus our Hamiltonian analysis will proceed from 
(\ref{hdlag}). Note that the higher order model was also considered in \cite{cordero1} where the Hamiltonian analysis was performed following the Ostrogradsky approach.
 We on the contrary follow the equivalent first order approach of \cite{BMP}.

\section{Hamiltonian analysis }
This section contains the main results of the present paper. As stated above our aim is to develop a new Hamiltonian analysis following from the Lagrangian (\ref{hdlag}) which is a second order theory. A Hamiltonian analysis of the same model has been discussed in \cite{cordero1} from the Ostrogradsky approach. We on the other hand adopt the equivalent first order formalism which has been demonstrated to be useful, specifically in treating the gauge invariances from the Hamiltonian point of view \cite{BMP, MP, paul}. The point of departure is to convert (\ref{hdlag}) to a first order theory by defining the first derivative of $a$ and $t$ as additional fields and including the following constraints into the Lagrangian with the help of undetermined multipliers. These multipliers are then treated as new fields and the phase space is constructed by the entire set of fields along with their conjugate momenta defined in the usual way as is done for first order theories. Automatically primary constraints arise. The constraint analysis is then presented in detail. In addition to first class constraints the model also has second class constraints. The second class constraints are then strongly implemented by substituting the Poisson brackets by the corresponding Dirac brackets. Effectively the theory becomes a first class system with the symplectic algebra given by these Dirac brackets of which a complete list has been given.

 The results derived so far are then used in two ways. First an analysis of the gauge invariances of the model has been done and its connection with the reparametrization invariance of the action has been discussed. Secondly, the gauge redundancy of the model has been eliminated by choosing an appropriate gauge. The final Dirac brackets have been used to reduce the phase space and indicate a formal quantization of the model.

  In the equivalent first order formalism, we define the new  fields as,  
\begin{eqnarray}
\nonumber
\dot{a} &=& A\\
\dot{t} &=& T,
\end{eqnarray}
which also introduce new constraints in the system given by 
 \begin{eqnarray}
\nonumber
A -\dot{a} \approx 0 \\
 T - \dot{t}  \approx 0,
\label{hdconst}
\end{eqnarray}
 Now the HD Lagrangian (\ref{hdlag}) is transformed to the first order Lagrangian  where the constraints (\ref{hdconst}) are enforced through the Lagrange multipliers $\lambda_{a}$, and $\lambda_{t}$ as 
\begin{equation}
L^{\prime}=\frac{aT}{\left(T^{2}-A^{2}\right)^{\frac{3}{2}}}\left( {aT\dot{A}-aA\dot{T}+\left(T^{2}-A^{2}\right)T}\right) - \left( T^{2}-A^{2}\right) ^{\frac{1}{2}}a^{3}H^{2} + \lambda_{a}\left(A-\dot{a}\right) + \lambda_{t}\left(T-\dot{t}\right).
\label{auxlag} 
\end{equation}
The Euler Lagrange equation of motion, obtained    from the first order Lagrangian (\ref{auxlag}),  by varying w.r.t. a, A, t, T, $\lambda_{a}$ and $\lambda_{t}$, are respectively given  by
\begin{eqnarray}
\frac{2a(\dot{A}T^{2} - AT\dot{T})}{(T^{2}-A^{2})^{\frac{3}{2}}} + \frac{T^{2}}{(T^{2}-A^{2})^{\frac{1}{2}}} -3 a^{2}H^{2}(T^{2}-A^{2})^{\frac{1}{2}} + \dot{\lambda}_{a} &=& 0
  \label{lag_eom_a} \\
 \frac{3a^{2}A(\dot{A}T^{2} - AT\dot{T})}{(T^{2}-A^{2})^{\frac{5}{2}}} - \frac{d}{d \tau}    \left( \frac{a^{2}T^{2}}{(T^{2}-A^{2})^{\frac{3}{2}}}  \right) - \frac{a^{2}T \dot{T}}{(T^{2}-A^{2})^{\frac{3}{2}}} + \frac{aAT^{2}}{(T^{2}-A^{2})^{\frac{3}{2}}} + \nonumber\\  \frac{a^{3}AH^{2}}{(T^{2}-A^{2})^{\frac{1}{2}}} + \lambda_{a} &=& 0 \label{lag_eom_A}\\ 
\dot{\lambda}_{t} &=& 0 \label{lag_eom_t} \\
\nonumber\\ \nonumber\\
\frac{3a^{2}T(\dot{A}T^{2} - AT\dot{T})}{(T^{2}-A^{2})^{\frac{5}{2}}} + \frac{2a^{2} \dot{A}T}{(T^{2}-A^{2})^{\frac{3}{2}}} - \frac{d}{d \tau} \left( \frac{a^{2} A T}{(T^{2}-A^{2})^{\frac{3}{2}}}\right) -\frac{a^{2} A \dot{T}}{(T^{2}-A^{2})^{\frac{3}{2}}} + \frac{2aT}{(T^{2}-A^{2})^{\frac{1}{2}}} \nonumber\\ - \frac{aT^{3}}{(T^{2}-A^{2})^{\frac{1}{2}}} + \lambda_{t} &=& 0 \label{lag_eom_T}\\
A-\dot{a}&=&0 \label{lag_eom_lambda_a}\\
T-\dot{t}&=&0. \label{lag_eom_lambda_t}
\end{eqnarray} 
Eliminating the multipliers $\lambda_{a}$, and $\lambda_{t}$ from the above equations we get back equation (\ref{ELeqn})

In the Hamiltonian formulation adopted in the present paper the Lagrange multipliers are considered formally as independent fields and the momenta corresponding to them are introduced in the usual way. Here we denote  the phase space coordinates by $q_{\mu}=a, t, A, T, \lambda_{a},\lambda_{t}$ and their corresponding momenta  as $\Pi_{q_\mu}=\Pi_{a},\Pi_{t}, \Pi_{A}, \Pi_{T}, \Pi_{\lambda_{a}}, \Pi_{\lambda_{t}}$  with $\mu=0, 1, 2, 3, 4, 5$. We adopt the usual definition 
\begin{equation}
\Pi_{q_{\mu}}  = \frac{\partial{L}^{\prime}}{\partial{\dot{q}_{\mu}}},
\end{equation} 
since the Lagrangian (\ref{auxlag}) is in the first order form. This is the point of departure of our Hamiltonian formulation from the Ostrogradsky formulation of \cite{cordero1}.
 
 From the definition of the phase space variables, we  get the following primary constraints
\begin{eqnarray}
\nonumber\\
\Phi_{1}&=& \Pi_{t} + \lambda_{t} \approx 0
\nonumber\\
\Phi_{2}&=& \Pi_{a} + \lambda_{a} \approx 0
\nonumber\\
\Phi_{3}&=&\Pi_{T} + \frac{a^{2}TA}{\left(T^{2}- A^{2}\right)^{\frac{3}{2}}} \approx 0
\nonumber\\
\Phi_{4}&=& \Pi_{A} - \frac{a^{2}T^{2}}{\left(T^{2}-A^{2}\right)^{\frac{3}{2}}} \approx 0
\nonumber\\
\Phi_{5}&=& \Pi_{\lambda_{t}}\approx0
\nonumber\\
\Phi_{6}&=& \Pi_{\lambda_{a}} \approx 0.
\end{eqnarray}

The nonzero Poisson brackets between the primary constraints are computed as
\begin{eqnarray}
\nonumber\\
\left\lbrace {\Phi_{1}, \Phi_{5}}\right\rbrace &=& 1
\nonumber\\
\left\lbrace {\Phi_{2}, \Phi_{3}}\right\rbrace &=& -\frac{2aTA}{\left(T^{2}-A^{2}\right)^{\frac{3}{2}}}
\nonumber\\
\left\lbrace {\Phi_{2}, \Phi_{4}}\right\rbrace &=& \frac{2aT^{2}}{\left(T^{2}-A^{2}\right)^{\frac{3}{2}}}
\nonumber\\
\left\lbrace {\Phi_{2}, \Phi_{6}}\right\rbrace &=& 1.
\end{eqnarray}
Taking the constraint combination  
\begin{equation}
\Phi_{3}^{\prime} = T\Phi_{3} + A\Phi_{4}, \approx 0,
\end{equation}
 we find that $\Phi_{3}^{\prime}$ commutes with all the constraints. The nonzero poisson brackets between the newly defined primary set of constraints $\Phi_{1},\Phi_{2},\Phi_{3}^{\prime},\Phi_{4},\Phi_{5},\Phi_{6},$ become
 \begin{eqnarray}
\nonumber\\
\left\lbrace {\Phi_{1}, \Phi_{5}}\right\rbrace &=& 1
\nonumber\\
\left\lbrace {\Phi_{2}, \Phi_{4}}\right\rbrace &=& \frac{2aT^{2}}{\left(T^{2}-A^{2}\right)^{\frac{3}{2}}}
\nonumber\\
\left\lbrace {\Phi_{2}, \Phi_{6}}\right\rbrace &=& 1.
\end{eqnarray}
We can write down the canonical Hamiltonian as  
\begin{eqnarray}
\nonumber\\
H_{can} &=& \Pi_{q_{\mu}}\dot{q}_{\mu} - L^{\prime} 
\nonumber\\
        &=& -\frac{aT^{2}}{\left(T^{2}-A^{2}\right)^{\frac{1}{2}}} + \left(T^{2}-A^{2}\right)^{\frac{1}{2}} a^{3}H^{2} -\lambda_{a}A - \lambda_{t}T.
\end{eqnarray}

 The total Hamiltonian is given by
 \begin{equation}
 H_{T} = H_{can} + \Lambda_{1}\Phi_{1} + \Lambda_{2}\Phi_{2} +  \Lambda_{3}\Phi_{3}^{\prime} + \Lambda_{4}\Phi_{4}+ \Lambda_{5}\Phi_{5} + \Lambda_{6}\Phi_{6}.
 \label{H_T} 
 \end{equation}
 Here $\Lambda_{1},\Lambda_{2}, \Lambda_{3}, \Lambda_{4}, \Lambda_{5}, \Lambda_{6}$ are  undetermined Lagrange multipliers.
  Preserving the primary constraints $\Phi_{1}$, $\Phi_{5}$, $\Phi_{6}$  in time ($ \{\Phi_{i}, H_{T}\} \approx 0$) the following Lagrange multipliers get fixed
 \begin{eqnarray}
 \nonumber\\
 \Lambda_{5}&=& 0 
 \nonumber\\
\Lambda_{1} &=& T
 \nonumber\\
\Lambda_{2} &=& A. 
 \nonumber
\end{eqnarray}
Whereas, conservation of $\Phi_{2}$ gives the following condition between $\Lambda_{4}$ and $\Lambda_{6}$
\begin{equation}
\frac{T^{2}}{\left(T^{2}-A^{2}\right)^{\frac{1}{2}}} - 3a^{2}H^{2}\left(T^{2} - A^{2}\right)^{\frac{1}{2}} + \Lambda_{6} + \Lambda_{4} \frac{2aT^{2}}{\left(T^{2}-A^{2}\right)^{\frac{3}{2}}}=0.
\label{32}
\end{equation}
 
Time preservation of the constraint  $\Phi_{3}^{\prime}$ gives rise to the following secondary constraint
 \begin{equation}
 \Psi_{1} = \frac{aT^{2}}{\left(T^{2}-A^{2}\right)^{\frac{1}{2}}} - a^{3}H^{2} \left(T^{2}-A^{2}\right)^{\frac{1}{2}} + \lambda_{t}T + \lambda_{a}A \approx 0.
\end{equation} 
Likewise, $\Phi_{4}$ yields the following secondary constraint 
\begin{equation}
\Psi_{2}=\frac{aAT^{2}}{\left(T^{2}-A^{2}\right)^{\frac{3}{2}}} - \frac{a^{3}H^{2}A}{\left(T^{2}-A^{2}\right)^{\frac{1}{2}}} - \lambda_{a} \approx 0.
\end{equation}
Nonzero brackets for $\Psi_{1}$ and $\Psi_{2}$  with the  other constraints are given below,
\begin{eqnarray}
\nonumber\\
\left\lbrace {\Phi_{2}, \Psi_{1}}\right\rbrace  &=& -\frac{T^{2} - 3a^{2}H^{2}\left(T^{2}-A^{2}\right)}{\left(T^{2}-A^{2}\right)^{\frac{1}{2}}}
\nonumber\\
\left\lbrace {\Phi_{4}, \Psi_{1}}\right\rbrace  &=& -\frac{aAT^{2}}{\left(T^{2}-A^{2}\right)^{\frac{3}{2}}} - \frac{a^{3}H^{2}A}{\left(T^{2}-A^{2}\right)^{\frac{1}{2}}} - \lambda_{a}
\nonumber\\
\left\lbrace {\Phi_{3}, \Psi_{1}}\right\rbrace  &=& - \frac{aT(2A^{2} - T^2)}{\left(T^{2}-A^{2}\right)^{\frac{3}{2}}} + \frac{a^{3} H^{2} T}{\left(T^{2}-A^{2}\right)^{\frac{1}{2}}} - \lambda_{t} 
\nonumber\\
\left\lbrace {\Phi_{5}, \Psi_{1}}\right\rbrace  &=& -T
\nonumber\\
\left\lbrace {\Phi_{6}, \Psi_{1}}\right\rbrace  &=& -A
\nonumber\\
\left\lbrace {\Phi_{2}, \Psi_{2}}\right\rbrace  &=& -\frac{AT^{2}}{\left(T^{2}-A^{2}\right)^{\frac{3}{2}}} + \frac{3a^{2}H^{2}A}{\left(T^{2} - A^{2}\right)^{\frac{1}{2}}}
\nonumber\\
\left\lbrace {\Phi_{4}, \Psi_{2}}\right\rbrace  &=& - \frac{aT^{2} \left(T^{2} +2A^{2}\right)}{\left(T^{2}-A^{2}\right)^{\frac{5}{2}}} + \frac{a^{3}H^{2}T^{2}}{\left(T^{2} - A^{2}\right)^{\frac{3}{2}}}
\nonumber\\
\left\lbrace {\Phi_{6}, \Psi_{2}}\right\rbrace  &=& 1.
\end{eqnarray}
Time preservation of  $\Psi_{1}$ trivially gives 0 = 0. A similar analysis involving $\Psi_{2}$ yields, on exploiting (\ref{32}),  

\begin{eqnarray}
\nonumber\\
\Lambda_{4} &=& - \frac{\left(T^{2} - 3a^{2}H^{2}\left(T^{2}-A^{2}\right)\right) \left(T^{2}-A^{2}\right)}{a\left(3T^{2} - a^{2}H^{2}\left(T^{2}-A^{2}\right)\right)}
\nonumber\\
\Lambda_{6} &=& - \frac{\left({ T^{2} - 3a^{2}H^{2}\left(T^{2}-A^{2}\right) }\right)  ( T^{2}-a^{2}H^{2}\left(T^{2}-A^{2}\right)^{\frac{1}{2}} )}{\left({T^{2}-A^{2}}\right)^{\frac{1}{2}} \left({3T^{2} -a^{2}H^{2} \left({T^{2}-A^{2}}\right)}\right)}.
\end{eqnarray}
The iterative procedure is thus closed and no more secondary constraints or other relations are generated.

 The above analysis reveals that of all the Lagrange multipliers $\Lambda_i$, only $\Lambda_1$ remains undetermined in (\ref{H_T}) signifying one independent gauge degree of freedom. This fact will be reflected in the gauge generator that has been constructed in section 3.1. It is interesting to note that this consistency is not always obvious in the Ostrogradsky formulation, as we have already mentioned in connection with the massive relativistic particle model  \cite{nesterenko}.

 We have now altogether eight primary and secondary constraints. Computation of the Poisson bracket between these constraints shows that only $\Phi_{3}^{\prime}$ is the first class constraint, whereas other seven  constraints are apparently second class. The odd number of apparently second class constraints signals the existence of additional first class constraints.  Indeed, 
the new constraint combination 
\begin{equation}
\Psi_{1}^{\prime} = \Psi_{1} - \Lambda_{1}\Phi_{1}- \Lambda_{2}\Phi_{2}- \Lambda_{4}\Phi_{4}- \Lambda_{5}\Phi_{5}- \Lambda_{6}\Phi_{6},
\label{new_combination}
\end{equation}
leads to a secondary first class constraint.
So now we have two  first class constraints $\Phi_{3}^{\prime}$, $\Psi_{1}^{\prime}$ and six  second class constraints $\Phi_{1}$, $\Phi_{2}$, $\Phi_{4}$, $\Phi_{5}$, $\Phi_{6}$ and $\Psi_{2}$. The total number of phase space variables is twelve. The number of independent phase space variables is therefore $12 - (2\times 2 +6)$ i.e. $2$. Later on we will explicitly identify these two variables. There is no enhancement of degrees of freedom as is customary for the higher derivative systems. This is consistent with the fact that (\ref{hdlag}) is not a genuine higher derivative system. Also, of the two first class constraints of the system,  $\Phi_{3}^{\prime}$ is the sole primary first class constraint. The number of primary first class constraint matches with the residual number of undetermined multiplier in the total Hamiltonian. This fact will be important in the construction of the gauge generator.

 To study gauge symmetry of the system we need to get rid of the second class constraints. This is done by the introduction of the Dirac brackets which enable us to  set these constraints strongly zero. For simplicity of the calculation we remove them pair by pair. The Dirac bracket between the basic fields after removing $\Phi_{1}, \Phi_{2}, \Phi_{5}, \Phi_{6}$ remains same as their corresponding Poisson brackets.    
Solving $\Phi_{1}, \Phi_{2}, \Phi_{5}, \Phi_{6}$ the new constraint structure becomes 
\begin{eqnarray}
\nonumber\\
F_{1}&=&\Phi_{3}^{\prime}  =T\Phi_{3} + A \Phi_{4} \approx 0
\nonumber\\
 F_{2}&=&\Psi_{1}^{\prime}= \Psi_{1} - \Lambda_{4}\Phi_{4} \approx 0
\nonumber\\
S_{1} &=&  \Phi_{4} \approx 0
\nonumber\\
S_{2}&=& \Psi_{2}= \frac{aAT^{2}}{\left({T^{2}-A^{2}}\right)^{\frac{3}{2}}} - \frac{a^{3}AH^{2}}{\left(T^{2}-A^{2}\right)^{\frac{1}{2}}} + \Pi_{a} \approx 0.
\label{secondclass}
\end{eqnarray}
For simplicity we use new notations \{$F_{1}$, $F_{2}$\} and  \{$S_{1}$, $S_{2}$\} where, the first pair denotes the set of first class constraint and second pair denotes the remaining  set of second class constraints. Some details of this reduction are given below.

To calculate Dirac brackets of the theory we first find out the Poisson brackets between the second class constraints which are written as 
\begin{equation}
\Delta_{ij}=\{S_{i}, S_{j}\}= -\frac {aT^{2} \left({3T^{2}-a^{2}H^{2} \left({T^{2} -A^{2}}\right)}\right)}{\left(T^{2}-A^{2}\right)^{\frac{5}{2}}}  \epsilon_{ij},
\end{equation}
with $\epsilon_{12} = 1$ and $ i, j = 1, 2$. Dirac brackets are defined by
\begin{equation}
\{f,g\}_{D} = \{f,g\} -\{f, S_{i}\}\Delta_{ij}^{-1}\{S_{j},g\}.
\label{dbdef}
\end{equation}
 We calculate the Dirac brackets between the basic fields which are given below(only the nonzero brackets are listed)
\begin{eqnarray}
\nonumber\\
\left\lbrace {a,A}\right\rbrace_{D}&=& -\frac {\left(T^{2}-A^{2}\right)^{\frac{5}{2}}}{aT^{2} \left({3T^{2}-a^{2}H^{2} \left({T^{2} -A^{2}}\right)}\right)}
\nonumber\\
\left\lbrace {a, \Pi_{a}}\right\rbrace_{D}&=& \frac{T^{2}+ 2 A^{2} - a^{2}H^{2} \left(T^{2}-A^{2}\right)}{\left(3T^{2} - a^{2}H^{2}\left(T^{2} - A^{2}\right) \right)}
 \nonumber\\
\left\lbrace {a, \Pi_{A}}\right\rbrace_{D}&=& -\frac{3aA}{3T^{2} - a^{2}H^{2} \left(T^{2}-A^{2}\right)}
 \nonumber\\
\left\lbrace {a, \Pi_{T}}\right\rbrace_{D}&=& \frac{a\left(T^{2}+ 2A^{2}\right)}{T\left(3T^{2}-a^{2}H^{2}(T^{2}-A^{2})\right)}
\nonumber\\
\left\lbrace {t,\Pi_{t}}\right\rbrace_{D}&=&1 \nonumber\\
\left\lbrace {A, \Pi_{a}}\right\rbrace_{D}&=& - \frac{A \left(T^{2}-A^{2}\right) \left(T^{2}-3a^{2}H^{2} \left(T^{2}-A^{2}\right)\right)}{aT^{2}\left(3T^{2} - a^{2}H^{2} \left(T^{2}-A^{2}\right)\right)} 
\nonumber\\
\left\lbrace {A, \Pi_{A}}\right\rbrace_{D}&=& \frac{2 \left(T^{2}-A^{2}\right)}{3T^{2} - a^{2}H^{2} \left(T^{2}-A^{2}\right)}
 \nonumber\\
 \left\lbrace{A, \Pi_{T}}\right\rbrace_{D}&=& \frac{A \left(T^{2} + 2A^{2} - a^{2}H^{2} \left(T^{2}-A^{2}\right) \right)}{T\left(3T^{2} - a^{2}H^{2} \left(T^{2}-A^{2}\right)\right)}
 \nonumber\\
\left\lbrace {T, \Pi_{T}}\right\rbrace_{D}&=& 1 \nonumber\\
\left\lbrace {\Pi_{a}, \Pi_{A}}\right\rbrace_{D}&=& -\frac{a \left(2T^{4}+A^{2}T^{2} + a^{2}H^{2}(T^{2}-A^{2})(9A^{2}-2T^{2})\right)}{ \left(T^{2}-A^{2}\right)^{\frac{3}{2}}(3T^{2} - a^{2}H^{2} \left(T^{2}-A^{2}\right)) }
 \nonumber\\
\left\lbrace {\Pi_{a}, \Pi_{T}}\right\rbrace_{D}&=& \frac{aA \left(T^{4}+2T^{2}A^{2} + a^{2}H^{2}(T^{2}-A^{2})(T^{2}+6A^{2})\right)}{T\left(T^{2}-A^{2}\right)^{\frac{3}{2}}(3T^{2} - a^{2}H^{2} \left(T^{2}-A^{2}\right) )}
 \nonumber\\
\left\lbrace {\Pi_{A}, \Pi_{T}}\right\rbrace_{D}&=& -\frac{a^{2}T \left(T^{2} +2A^{2} - a^{2}H^{2} (T^{2}-A^{2})\right)}{\left(T^{2}-A^{2}\right)^{\frac{3}{2}}(3T^{2} - a^{2}H^{2} \left(T^{2}-A^{2}\right))}.
\label{intdb} 
\end{eqnarray}

 The introduction of the above Dirac brackets allows the second class pair \{$S_{1}$, $S_{2}$\} to be strongly implemented. Note that the secondary first class constraint then becomes equal to the canonical Hamiltonian: 
\begin{equation}
F_{2}=\Psi_{1}= -H_{c}=-T\Pi_t - \frac{T^2}{A}\Pi_a \approx 0.
\end{equation} 
Vanishing of the canonical Hamiltonian is a consequence of the reparametrisation invariance of the theory. 

 

\subsection{Construction of the gauge generator}
The equivalent first order formalism offers a structured algorithm for the abstraction of the gauge generator of the higher derivative system \cite{BMP} which is based on the method presented in \cite{BRR1, BRR2} for the first order systems.
According to the Dirac conjecture \cite{dirac} the gauge generator is 
\begin{equation}
G = \sum_a \epsilon_a \Phi_a.
\label{217}
\end{equation}
 Here $\{\Phi_a\}$ is the whole set of  constraints and $\epsilon_{a}$ are the gauge parameters. However not all the gauge parameters $\epsilon_{a}$ are independent. The number of independent gauge parameters is equal to the number of independent primary first class constraints \cite{BRR1, BRR2} . Demanding the commutativity of gauge variation and time translation 
we get the following master equations
  \begin{equation}
\delta\Lambda_{a_{1}} = \frac{d\epsilon_{a_{1}}}{dt}
                 -\epsilon_{a}\left( {V_{a a_{1}}
                 +\Lambda_{b_{1}}C_{b_{1} a a_{1}} }\right)                            
                           \label{master1}
\end{equation}
\begin{equation}
  0 = \frac{d\epsilon_{a_{2}}}{dt}
 -\epsilon_{a}\left(V_{a a_2}
+\Lambda_{b_1} C_{b_1 a a_2}\right).
\label{master2}
\end{equation}
 Here the indices $a_1, b_1 ...$ refer to the primary first class constraints while the indices $a_2, b_2 ...$ correspond to the secondary first class constraints. $\Lambda_{a_{1}}$ are the Lagrange multipliers multiplying the primary first class constraints in the expression of the total Hamiltonian and $\delta$ denotes gauge variation.
The coefficients $V_{aa_{1}}$ and $C_{b_1aa_1}$ etc. are the structure functions of the involutive algebra, defined as \footnote{from now on we have to use only Dirac brackets since we removed all second class constraints. Poissson brackets are denoted by $\{ \  , \}$  , whereas, $\{ \  , \}_{D}$ refers to Dirac brackets}
\begin{eqnarray}
\{H_{can},\Phi_{a}\}_{D} = V_{ab}\Phi_{b}\nonumber\\
\{\Phi_{a},\Phi_{b}\}_{D} = C_{abc}\Phi_{c}.
\label{2110}
\end{eqnarray}
Equations (\ref{master1}) give no new conditions as they can be shown to follow from  (\ref{master2}) \cite{BRR1}. The latter equations actually impose restrictions on the gauge parameters. Using these  the independent gauge parameters can be identified. A new feature appears in case of the HD theories where in the equivalent first order formalism we define the time  derivatives of the coordinates right upto one order less than the highest order appearing in the Lagrangian as independent fields. Thus the gauge variations here must be consistent with this definition and we require conditions of the form
\begin{eqnarray}
\delta q_{n,\alpha} - \frac{d}{dt}\delta{q}_{n,\alpha -1} = 0, \left(\alpha > 1 \right),
\label{varsgauge}
\end{eqnarray} 
where $q_{n,\alpha}$ denotes the $\alpha$-th order time derivative of $q$.
The conditions (\ref{varsgauge}) sometimes impose some extra condition on the gauge parameters and sometimes not \cite{BMP,MP,paul}.
Expressing the gauge parameters in terms of the independent elements of the set in (\ref{217}) the most general form of the gauge generator is constructed. Now we can write gauge variations of the basic fields as
\begin{equation}
\delta_{\epsilon_{a}} {q_{n,\alpha}} = \{ q_{n,\alpha}, G \}_{D}.
\end{equation}
 where on the right hand side only the independent gauge parameters appear.
  
 After the short review of the basic methodology we come back to the present model. The gauge generator is defined as the linear combination of all the first class constraints which is written as,
\begin{equation}
 G= \epsilon_{1} F_{1} + \epsilon_{2} F_{2}.
 \label{gaugegenerator11}
 \end{equation}
 Here $\epsilon_{1}$ and $\epsilon_{2}$ are the gauge parameters. 
From equations (\ref{2110}) we find that $C_{122} = -1 = -C_{212}$ and $V_{12}=1$  are the only nonzero structure functions. Now using equation (\ref{master2})the following relation between the gauge parameters is obtained 
 \begin{equation}
 \epsilon_{1}= - \Lambda_{3}\epsilon_{2} - \dot{\epsilon}_{2}. 
 \label{parameterrelation1}
 \end{equation}
 So here  $\epsilon_{2}$ may be chosen as the independent gauge parameter. 

 At this stage we observe that there is one independent parameter in the gauge generator (\ref{gaugegenerator11}). The conditions (\ref{varsgauge}) following from the higher derivative nature is yet to be implemented. As has been mentioned earlier this may or may not impose additional restriction on the gauge parameters. 
 The gauge transformations of  the fields  are given by 
 \begin{eqnarray}
 \delta{a} &=& \{a, G\}_D = -  \epsilon_{2} A \label{gaugetrans_a}\\
 \delta{t} &=& -  \epsilon_{2} T \label{gaugetrans_t}\\
 \delta{A} &=&   \epsilon_{1} A  - \epsilon_{2} \frac{\left(T^{2} - 3a^{2}H^{2} \left(T^{2}-A^{2}\right)\right)\left(T^{2}-A^{2}\right)}{a\left(3T^{2} - a^{2}H^{2} \left(T^{2}-A^{2}\right)\right)}
  \label{gaugetrans_A}\\
 \delta{T} &=&   \epsilon_{1} T 
 \label{gaugetrans_T}
 \end{eqnarray}
After some calculation we find that
 \begin{eqnarray}
 \nonumber\\
 \frac{d}{d \tau}\delta {a} &=& \delta A \label{hd_gauge_a}\\
   \frac{d}{d \tau}\delta {t} &=& \delta T. \label{hd_gauge_t}
 \end{eqnarray}
So the constraints (\ref{varsgauge}) hold identically for the present model and impose no new condition on the gauge parameters. We find therefore that there is only one independent gauge transformation which essentially is in conformation with the fact that there is only one independent primary first class constraint.
 
 The gauge variations obtained from the Hamiltonian analysis can be exactly mapped to the reparametrization invariance of the model. Consider arbitrary infinitesimal  change in the parameter $\tau \rightarrow \tau^{\prime} = \tau + \sigma$. The action is invariant under this reparametrization. Now the fields transform as
\begin{eqnarray}
\nonumber
\delta a &=&  - \sigma a \\
\delta t &=&  - \sigma t.
\end{eqnarray} 
These are identical with the gauge variations (\ref{gaugetrans_a}) and (\ref{gaugetrans_t}) of $a$ and $t$ if $\sigma$ is identified with $\epsilon_{2}$. The equivalence of gauge invariances with the reparametrization invariance of the model is thus established.

\subsection{Gauge fixing and formal quantization}
 After the reduction of phase space by the  Dirac bracket procedure we are left with only the two first class constraints $F_1$ and $F_2$. These first class constraints reflect the redundancy of the theory which are connected by gauge transformations. In the above analysis our focus was on the abstraction of the gauge degrees of freedom. We now elucidate a formal quantisation prescription. A gauge fixing is done and the appropriate WDW equation is written.

           The choice of gauge is arbitrary subject to the conditions that they must reduce the first class constraints to second class. Also the constraint algebra should be nonsingular. As there are two first class constraints we need two gauge conditions. We take one of these to be the cosmic gauge
\begin{eqnarray}
\varphi_{1} &=& \sqrt{T^2 - A^2} -1 \approx 0.
\label{gauge1} 
\end{eqnarray}
The name derives from the fact that the resultant metric becomes the usual FLRW metric. As the second gauge condition we take  
\begin{eqnarray}
\varphi_{2} &=& T - \alpha a \approx 0.
\label{gauge2}
\end{eqnarray}
where the constant $\alpha$ is chosen so that $\alpha \ne H$. The following calculations will show that these are appropriate gauge conditions.

           As usual the gauge conditions are treated as additional constraints which make the first class constraints of the theory second class. For convenience, renaming the two first class constraints we write the complete set of constraints as
\begin{eqnarray}
\Omega_{1} = F_{1}\\
\Omega_{2} = F_{2}\\
\Omega_{3} = \varphi_{1}\\ 
\Omega_{4} =  \varphi_{2}.
\end{eqnarray}
Modifying the algebra by the Dirac brackets corresponding to this second class system we will be able to put  all the second class constraints ($\Omega_i, i = 1, 2, 3, 4$) to be strongly equal to zero. These will correspond to operator relations in the corresponding quantum theory.

        Using the algebra (\ref{intdb}) we can straightforwardly compute the algebra of the constraints $\Omega_i$. The results are given in the following table
 \begin{table}[ht]
\caption{Constraint brackets} 
\centering 
\begin{tabular}{c c c c c} 
\hline\hline
 & $\Omega_{1}$ & $\Omega_{2}$& $\Omega_{3}$ & $\Omega_{4}$ \\ [1.0 ex] 
\hline 

$\Omega_{1}$ & 0 & 0 & -1 & $-T$  \\
$\Omega_{2}$ & 0 &0 & $\frac{A (\alpha^{2} - 3 H^2)}{a (3\alpha^{2} - H^2)}$ & $- \alpha A$ \\
$\Omega_{3}$ & 1 & $-\frac{A (\alpha^{2} - 3 H^2)}{a (3\alpha^{2} - H^2)}$ & 0 & $\frac{A}{\alpha a^5 (3 \alpha^2 - H^2)}$\\
$\Omega_{4}$ & $T$ & $ \alpha A$&   $-\frac{A}{\alpha a^5 (3 \alpha^2 - H^2)}$ & 0 \\ [1ex] 
\hline 
\end{tabular}
\label{table:nonlin} 
\end{table}\\
From the above table we can read off the matrix
\begin{eqnarray}
\nonumber\\
\Delta_{ij} = \{\Omega_i,\Omega_j\}
\end{eqnarray}
Using the definition (\ref{dbdef}) we can calculate the final Dirac brackets.
Nonzero Dirac brackets between the phase space variables are
\begin{eqnarray}
\nonumber\\
\{ t , a\}^{*} &=& \frac{1}{4 \alpha a^3 (\alpha^2 - H^2)}
\nonumber\\
\{ t , A \}^{*} &=& \frac{\alpha}{4  a^2 A (\alpha^2 - H^2)}
\nonumber\\
\{ t , T \}^{*} &=& \frac{1}{4 a^3 (\alpha^2 - H^2)}
\nonumber\\
\{ t , \Pi_{a}\}^{*} &=& \frac{-4 a^2 \alpha^2 + 3}{4 \alpha a A}
\nonumber\\
\{ t , \Pi_{A} \}^{*} &=& \frac{\alpha}{(\alpha^2 - H^2)}
\nonumber\\
\{ t , \Pi_{t}\}^{*} &=& 1
\nonumber\\
\{ t , \Pi_{T}\}^{*} &=& \frac{-4 a^2 \alpha^2 + 3}{4  a A (\alpha^2 - H^2)}.
\label{final_dirac}
\end{eqnarray}
With the introduction of the final Dirac brackets all the constraints (including the gauge conditions) become second class
and strongly zero. We thus have the following conditions on the phase space variables 
\begin{eqnarray}
\Pi_{T} + \frac{a^{2}TA}{\left(T^{2}- A^{2}\right)^{\frac{3}{2}}} &=& 0
\nonumber\\
 \Pi_{A} - \frac{a^{2}T^{2}}{\left(T^{2}-A^{2}\right)^{\frac{3}{2}}} &=& 0
 \nonumber\\
 -\Pi_{t} - \frac{T}{A}\Pi_{a} &=& 0
 \nonumber\\
 \frac{aAT^{2}}{\left({T^{2}-A^{2}}\right)^{\frac{3}{2}}} - \frac{a^{2}AH^{2}}{\left(T^{2}-A^{2}\right)^{\frac{1}{2}}} + \Pi_{a} &=& 0
 \nonumber\\
 \sqrt{T^2 - A^2} &=& 1
 \nonumber\\
 T - \alpha a &=& 0
\end{eqnarray}
 where use has been made of equations (\ref{secondclass}, \ref{gauge1},   \ref{gauge2}). From the final Dirac brackets (\ref{final_dirac}) it is clear that only the pair $(t, \Pi_t)$ is canonical. We thus identify this pair as the two independent phase space degrees of freedom found earlier by a standard count using the constraints of the system (see below {\ref{new_combination}}). To develop a quantum theory  it is necessary to write down the whole theory with respect to the canonical variables in the reduced phase. All the variables can be expressed in favour of $(t, \Pi_t)$ by appropriately solving the constraints which are now strongly implemented. The result is,
\begin{eqnarray}
\nonumber\\
T &=& \alpha a \nonumber\\
A &=& \sqrt{\alpha^{2} a^2 - 1} \nonumber\\
\Pi_{A} &=& \alpha^{2} a^{4} 
\nonumber\\
\Pi_{T} &=& - \alpha a^{3} \sqrt{\alpha^{2} a^{2} - 1} \nonumber\\
\Pi_{a} &=&  - a^{3}(\alpha^{2} - H^2) \sqrt{\alpha^{2} a^2 - 1}.
\label{gaugephase1} 
\end{eqnarray}
where $a$ is expressed as
\begin{equation}
a = \left( {\frac{\Pi_{t}}{\alpha(\alpha^2 - H^2)}} \right) ^{\frac{1}{4}}
\label{gaugephase2}.
\end{equation}
Thus we find that all the phase space variables except $t$ are given as function of $\Pi_{t}$. 

The passage from the classical to quantum theory proceeds in the usual way. The phase space variables are lifted to operators in some Hilbert space and the conditions (\ref{gaugephase1}, \ref{gaugephase2}) are now treated as operator relations. The Dirac brackets are promoted to commutators according to the prescription.  
\begin{equation}
\{B,C\}^* \to \frac{1}{i\hbar}[B,C].
\end{equation}
The fundamental canonical algebra is thus (with $\hbar = 1$ )
\begin{equation}
[t, t] =[\Pi_{t}, \Pi_{t}] = 0, \ \   [t, \Pi_{t}] = i. 
\label{commutator}
\end{equation}
	
	We next proceed to formulate the WDW equation for the universe governed by the Lagrangian (\ref{orglag}). Before that we  write down the first class constraint $F_2$ which is the canonical Hamiltonian as
\begin{equation}
F_2 = -H_{can} = - \frac{-A^{2}\Pi_{t}^{2}+T^{2}\Pi_{a}^{2}}{A\Pi_{a}} = 0
\label{final_canonical}
\end{equation}
 Considering the sate vector $|\Psi\rangle$ in the appropriate Hilbert space, the WDW equation may be written as,
 \begin{equation}
 H_{can} |\Psi\rangle=0.
 \end{equation}
 Using the Schrodinger representation compatible with (\ref{commutator}), we obtain,
 \begin{equation}
 \Pi_t = -i\frac{\partial}{\partial{t}}
 \label{schrodinger}
 \end{equation}
 Exploiting (\ref{final_canonical}-\ref{schrodinger}) and the expression for $A$ given in (\ref{gaugephase1}) we obtain, after some algebra, the following WDW equation,
 \begin{equation}
-\frac{\partial^{2}}{\partial{t^2}} |\Psi\rangle = \alpha^2 a^8 (\alpha^2 - H^2)^2 |\Psi\rangle.
\label{WDW1}
 \end{equation}
 Making a change of variables $\xi=\frac{\alpha^2}{H^2}$, the WDW equation may be reexpressed as,
 \begin{equation}
 -\frac{\partial^{2}}{\partial{t^2}} |\Psi\rangle =  \xi (\xi - 1)^{2}H^{6}a^{8}  |\Psi\rangle.
\label{WDW2}
\end{equation}  
The above equation exactly reproduces one piece of the bifurcated WDW equation found in the first item of \cite{davidson1} \footnote{Note that the other part of the bifurcated WDW involving the $`a'$ variable is nonexistent in the present analysis. This is because here we have only one (configuration space) independent degree of freedom (i.e. $t$) instead of two variables ($t$ and $a$) as occurs in \cite{davidson1}. This mismatch happens because, contrary to \cite{davidson1}, the present analysis is done in a fully reduced space where all constraints are eliminated} .

	Furthermore, introducing the conserved `energy' $\omega$ by,
	\begin{equation}
	 \xi (\xi - 1)^{2}H^{6}a^{8} = \omega^2
	 \label{conserved_energy}
	\end{equation}
	we may reexpress (\ref{WDW2}) by the standard equation,
	\begin{equation}
 -\frac{\partial^{2}}{\partial{t^2}} |\Psi\rangle =  \omega^2  |\Psi\rangle.
\label{WDW3}
\end{equation} 
The expression for the conserved energy $\omega$ in (\ref{conserved_energy}) matches with the form given in \cite{davidson1}. It is now possible to proceed with the quantisation as elaborated in \cite{davidson1}.

	Before concluding this section it is worthwhile to mention the efficacy of the gauge choice (\ref{gauge2}). While the first gauge condition (\ref{gauge1}) is the standard cosmic gauge, the second one (\ref{gauge2}) has not been considered earlier. We have shown that this simple choice (\ref{gauge2}) is a valid choice that yields the fully reduced space of the model. Also, at the quantum level, the WDW equation subjected  to this gauge fixing reproduces the expression obtained earlier in \cite{davidson1}.  
 \section{Conclusions}

  The minisuperspace cosmology following from the geodetic brane gravity model introduced  by Regge and Teitelboim \cite{regge} has been considered from the point of view of higher derivative theory following Cordero, Molgado and Rojas \cite{cordero1}. We have presented a new Hamiltonian formulation of the model based on the equivalent first order formalism
\cite{BMP, MP, paul}. This is different from the analysis of \cite{cordero1} where the usual Ostrogradsky approach is adopted. Not only that our equivalent first order formalism differs from the first order Hamiltonian formalism for the model obtained by dropping a boundary term from the action \cite{karasik}. The latter is plagued with problems that can be eradicated only by the introduction of an auxiliary field. Our Hamiltonian formalism is free from such difficulties. Apart from this the present equivalent first order approach is known to provide greater control in treating singular systems as has recently been demonstrated in connection with the massive relativistic model with curvature term \cite{BMP}. Specifically, an analysis of the later model from the Ostrogradsky approach \cite{nesterenko} yields two primary first class constraints whereas the total Hamiltonian contains only one arbitrary multiplier signifying only one gauge degree of freedom. Thus the number of primary first class constraints does not match the number of gauge degrees of freedom as happens in usual first order systems. This paradox was resolved in \cite{BMP} using the equivalent first order approach where a well known algorithm for constructing the Hamiltonian gauge generator \cite{BRR1, BRR2} was used along with conditions imposed due to the higher derivative nature. This additional constraint may or may not lead to an independent restriction on the gauge generator \cite{BMP, MP, paul}. It did impose an independent additional restriction on the gauge invariances of the massive relativistic model with curvature which explained the apparent mismatch between the number of primary first class constraints and the number of independent gauge degrees of freedom mentioned above \cite{BMP}.

            We have provided a complete Hamiltonian analysis of the minisuperspace Regge-Teitelboim cosmological model using the equivalent first order approach. The model was treated as a second order theory. The first derivatives of the fields have been defined as new coordinates. This redefinition  led to Lagrangian constraints. The original Lagrangian of the model was then converted to an equivalent first order Lagrangian by incorporating the constraints by the Lagrange multiplier technique. These multipliers were considered as independent fields in the Hamiltonian analysis where their conjugate momenta have been introduced in the usual way as is done for the first order systems. The full constraint structure has been worked out. The second class constraints of the model were then strongly implemented by  substituting the Poisson brackets by the corresponding Dirac brackets. 

            The results of the Hamiltonian analysis detailed above have been used in two ways. First we construct the  gauge generator using the algorithm of \cite{BRR1,BRR2}. For convenience a short review of this algorithm is provided. The gauge generator is first constructed as a linear combination of {\it{all}} the first class constraints of the theory. The structure functions are worked out from the algebra of the first class constraints with respect  to the Dirac brackets referred above. These structure functions are plugged in the master equation connecting the gauge parameters provided by the chosen algorithm. One relation is found between the two gauge parameters appearing in the gauge generator.
The additional constraints following from the higher derivative nature were shown to hold identically. Thus only one gauge parameter was found to be independent. There was only one primary first class constraint. So in this case the number of independent gauge parameters was found to be equal to the number of primary first class constraints. Exact mapping of the Hamiltonian gauge invariances with the Lagrangian (reparametrization) invariances of the model has also been demonstrated. 

     The canonical quantization of the model is discussed next. For this the redundancy of the phase space  was eliminated by choosing appropriate gauge fixing conditions. The familiar cosmic gauge was chosen as one of the gauge conditions. But the second gauge was a new one different from the nonstandard gauge chosen in \cite{cordero1}. As subsequent analysis revealed this new gauge condition is a good choice. A detailed account of the complete  gauge fixed calculations for the model has been presented. Formal quantization is obtained by promoting the phase space variables to operators in an assumed Hilbert space. The phase space is reduced so that only two phase space variables remain independent; the number being equal to the number of degrees of freedom in phase space. The fundamental commutator is then obtained from the Dirac bracket between the varibles according to well known procedure \cite{hanson}. The WDW equation which defines the quantum states of the universe corresponding to  the Lagrangian is constructed. This equation and the energy expression are shown to match with the existing literature \cite{davidson1}. Finally, we would like to mention a recent paper \cite{cordero2} where somewhat conclusions were obtained in the model considered here (\ref{main_lag}) augmented by an extrinsic curvature term.

 \section*{Acknowledgement}
 One of the authors  (BP) gratefully acknowledges Claus Kiefer for discussions. He also    acknowledges the Council of Scientific and Industrial Research (CSIR), Government of India, for financial assistance.

\end{document}